\documentclass[journal]{IEEEtran}
\IEEEoverridecommandlockouts
\usepackage{cite}
\usepackage{amsmath,amssymb,amsfonts,mathtools,bm}
\usepackage{amsthm}
\usepackage{algorithm}
\usepackage{algorithmic}
\usepackage{graphicx}
\usepackage{textcomp}
\usepackage{xcolor,float}
\usepackage{tabularx}
\usepackage{textcomp}
\usepackage{diagbox}
\usepackage{svg}
\usepackage{booktabs}
\usepackage{subfigure}
\usepackage{hyperref}
\usepackage{changes}

\newtheorem{lemma}{Lemma}
\newtheorem{problem}{Problem}
\usepackage{amssymb}
\usepackage{pifont}

\makeatletter

\newcommand{\Rmnum}[1]{\expandafter\@slowromancap\romannumeral #1@}
\makeatother

\allowdisplaybreaks
\renewcommand{\baselinestretch}{1}

\begin{document}

\title{Differentiable Projection-based Learn to Optimize in Wireless Network-Part I: Convex Constrained (Non-)Convex Programming}
\author{
Xiucheng Wang,~\IEEEmembership{Graduate Student Member,~IEEE,}
Xuan Zhao,~\IEEEmembership{Student Member,~IEEE,}\\
Nan Cheng,~\IEEEmembership{Senior Member,~IEEE}

\thanks{ }
\thanks{
\par Xiucheng Wang, Nan Cheng are with the State Key Laboratory of ISN and School of Telecommunications Engineering, Xidian University, Xi’an 710071, China (e-mail: xcwang\_1@stu.xidian.edu.cn; dr.nan.cheng@ieee.org). \textit{Nan Cheng is the corresponding author}.
\par Xuan Zhao is with the School of Cyber Engineering, Xidian University, Xi'an, 710071, China (e-mail: zhaoxuan@stu.xidian.edu.cn).
}

}

    \maketitle

\IEEEdisplaynontitleabstractindextext

\IEEEpeerreviewmaketitle

\begin{abstract}
This paper addresses a class of (non-)convex optimization problems subject to general convex constraints, which pose significant challenges for traditional methods due to their inherent non-convexity and diversity. Conventional convex optimization-based solvers often struggle to efficiently handle these problems in their most general form. While neural network (NN)-based approaches offer a promising alternative, ensuring the feasibility of NN-generated solutions and effectively training the NN remain key hurdles, largely because finite-capacity networks can produce infeasible outputs. To overcome these issues, we propose a projection-based method that projects any infeasible NN output onto the feasible domain, thus guaranteeing strict adherence to the constraints without compromising the NN's optimization capability. Furthermore, we derive the objective function values for both the raw NN outputs and their projected counterparts, along with the gradients of these values with respect to the NN parameters. This derivation enables label-free (unsupervised) training, reducing reliance on labeled data and improving scalability. Experimental results demonstrate that the proposed projection-based method consistently ensures feasibility.
\end{abstract}


\section{Introduction}
Constrained optimization (CO) is a fundamental approach in mathematical programming, wherein the objective is to obtain a minimum or maximum solution within a specified feasible region \cite{boyd2004convex}.\footnote{In this paper, only differentiable objective functions are considered.} Such formulations are extensively employed in wireless network management, particularly when resources are limited or when users’ personal preferences must be satisfied \cite{wang2022joint,9252917}. Consequently, the optimality of CO solutions significantly impacts the performance of wireless networks. Traditionally, convex programming with convex constraints can be solved effectively using well-established convex optimization algorithms, such as the interior point method.\footnote{Certain convex programs can still be non-deterministic polynomial (NP)-hard, which are beyond the scope of this paper \cite{de2002approximation}.} However, modeling wireless network problems with convex formulations remains challenging \cite{sun2016majorization}. For instance, although the optimization of the signal-to-noise ratio (SNR) can be formulated as a convex problem, the optimization of the signal-to-interference-plus-noise ratio (SINR)—widely employed in multicast and multi-transmitter scenarios—is non-convex \cite{9252917}. Furthermore, the complexity of SINR optimization increases dramatically as the number of antennas grows, particularly in large-scale and extremely large-scale multiple-input multiple-output (MIMO) systems, which constitute key technologies for beyond 5G (B5G) and 6G networks. Additionally, many 6G integrated sensing, communication, and computation (ISCC) tasks involve non-convex factorization problems, posing formidable challenges for ISCC optimization \cite{wen2024survey}. As a result, designing effective algorithms to tackle constrained non-convex problems is imperative for enhancing the performance of wireless networks.
\begin{table*}[ht]
    \centering
    \caption{Comparison of Constrained (Non-)Convex Programming Approaches}
    \resizebox{0.95\linewidth}{!}{
    \begin{tabular}{c|cccc}
    \hline
       Method  & Solution Feasibility Guarantee & Low Run-Time Complexity & Low Training Complexity & High Generalizability \\ \hline
       MM & \ding{51} & \ding{56} & N/A & \ding{51}\\
       AM & \ding{51} & \ding{56} & N/A & \ding{51}\\
       Penalty Loss Method & \ding{56} & \ding{51} & \ding{56} & \ding{56}\\
       \textbf{Projection-based L2O} (Ours) & \ding{51} & \ding{51} & \ding{51} & \ding{51}\\\hline
       \end{tabular}
    }
    \label{tab-method}
\end{table*}
Since most non-convex optimization problems are NP-hard, there is no universal method to solve them efficiently \cite{boyd2004convex}. A common strategy is to transform a non-convex problem into an equivalent convex form amenable to standard convex optimization techniques. Among these methods, majorization–minimization (MM) is one of the most popular, where a convex function is used to approximate the original non-convex function \cite{sun2016majorization}. The main advantage of MM is its ability to find a Karush–Kuhn–Tucker (KKT) point within the feasible region; however, constructing an appropriate majorization is often challenging. Although a second-order Taylor expansion can theoretically majorize a non-convex function with a quadratic approximation, its practical performance and computational efficiency can be poor, especially without a suitable initial point. To address specific types of non-convex problems, more specialized algorithms have been developed. For instance, if the objective function can be expressed as the difference of two convex functions, the problem can be cast as difference-of-convex (DC) programming, where a Fenchel conjugate function is iteratively minimized to identify a local optimum \cite{an2005dc}. The main drawback of DC lies in expressing a non-convex function as a difference of convex functions, which severely limits its applicability. Another widely employed non-convex approach is the alternating minimization (AM) scheme, which partitions the optimization variables into subsets so that each subset forms a convex subproblem \cite{niesen2007adaptive}. Although AM is frequently used in ISCC scenarios, it becomes ineffective if even a single subset remains non-convex (e.g., in SINR optimization) \cite{wen2024survey}. Similar benefits and limitations also arise in alternating direction optimization methods, such as the alternating direction method of multipliers (ADMM) \cite{nishihara2015general}.

In brief, employing traditional convex analysis methods to tackle non-convex problems typically results in high analytical complexity, low computational efficiency, and limited performance and generalizability. Because most existing non-convex algorithms are designed for specific problem forms and do not provide universal performance guarantees, they cannot be readily applied to arbitrary non-convex problems. Moreover, specialized methods such as DC, AM, and ADMM have narrow applicability, and identifying a suitable majorization can be difficult. Consequently, researchers have turned to neural networks (NNs) with universal approximation capabilities to develop learning-to-optimize (L2O) techniques, in which the NN learns to map problem parameters to the optimal solution. As a pioneering example, \cite{8444648} proposed a multilayer perceptron (MLP)-based L2O method, where channel state information (CSI) is provided as input to generate an optimized beamforming vector. This MLP is trained in a supervised manner, with labels derived from the weighted minimum mean-square error (WMMSE) solution. Although this approach substantially reduces computational complexity, its performance is constrained by the quality of the label-generation process. To overcome this limitation, PcNet \cite{8922744} introduced an unsupervised training approach, directly employing the objective function as the loss, thereby enabling PcNet to surpass WMMSE on average. As a further milestone, \cite{9252917} utilized a message-passing graph neural network (MPGNN) with permutation equivariance to enhance L2O performance, outperforming WMMSE-100—a method previously regarded as near-optimal.

The success of earlier L2O methods and scaling laws in computer science (CS) has led to the notion that increasingly large and complex neural networks (NNs) can effectively solve non-convex wireless network problems, whether constrained or unconstrained. However, the universal approximation theorem for NNs acts as a double-edged sword: although it ensures that a sufficiently large NN can approximate a broad class of functions, any practical NN with finite width inevitably exhibits non-negligible prediction errors \cite{goodfellow2016deep}. Consequently, there is always a certain probability that an NN will generate outputs of inappropriate magnitude, posing significant challenges for constrained optimization (CO). Moreover, merely enlarging an NN’s architecture does not guarantee it will respect simple, common-sense constraints, as demonstrated by large-scale generative models, even the generative pre-trained transformer (GPT)-4o \cite{achiam2023gpt} sometimes producing outputs with evident errors. In response, some researchers have drawn inspiration from classical optimization techniques, such as the Lagrange dual function and exterior penalty functions, to incorporate penalty terms for constraint violations directly into the NN’s loss function. Ideally, assigning higher penalties to infeasible solutions encourages the NN to remain within the feasible domain. Although this method can theoretically ensure convergence with probability one, it faces practical obstacles. First, NNs are usually designed to learn a fixed loss function, and altering that loss or its hyperparameters often triggers so-called “generalization problems.” Specifically, choosing a penalty coefficient that is too low may fail to guide the network toward the feasible domain, whereas an excessively high coefficient increases the risk of gradient explosion, hindering effective training. Additionally, incrementally adjusting the penalty term—akin to traditional outer-point penalty approaches—is restricted by the existence of a critical period in NN training, during which retraining with a different loss function becomes difficult. These issues make it challenging to guarantee convergence to the feasible region with probability one. To address these limitations, certain researchers have pursued specialized network architectures inherently suited to particular types of CO. One representative example is [X], in which a graph neural network (GNN) was crafted to solve linear equations directly, ensuring that the network’s output naturally satisfies linear constraints. However, this approach features high design complexity and demonstrates limited generalization when extended to nonlinear constraints.

A primary challenge in leveraging L2O for CO lies in designing a general framework capable of handling a wide variety of constrained optimization problems while ensuring that the NN’s outputs remain strictly within the feasible region. Although this task may appear daunting, it primarily stems from a mismatch between the fundamental properties of these CO problems and the conventional strategies employed to solve them using NNs. Most existing studies rarely address how to restrict NN outputs through activation functions. Nevertheless, an examination of prior L2O-based work in wireless networks reveals that these methods effectively tackle CO problems, though typically with simpler constraints—such as resource allocation or beamforming subject to total transmit power or bandwidth limits—where the optimization variables (or their sum) are constrained to be smaller than a fixed value and non-negative. In these cases, widely used activation functions such as softmax and sigmoid can naturally enforce non-negativity and limit the output range. However, for more intricate constraints, even if the feasible domain is a convex cube or a second-order cone, existing activation functions struggle to ensure that the NN’s outputs remain within valid bounds. Consequently, a more versatile activation function is needed to solve optimization problems involving more complex constraint sets. Designing such an activation function is not trivial. Beyond being generalizable, it must remain differentiable to avoid gradient vanishing issues that hamper NN training. Additionally, its derivative should be single-valued across its domain, thereby preventing a single input from mapping to multiple gradient directions, which can impede convergence. Crucially, the activation function’s output range must not be more restrictive than the feasible domain of the original CO problem; otherwise, achieving the true optimum becomes theoretically impossible.

Inspired by the Hilbert projection theorem, which points that the projection of any point in a space onto a closed convex set is unique. Moreover, determining this projection can be viewed as solving a second-order convex optimization problem, which can be handled efficiently using interior point methods or similar techniques. Consequently, the Hilbert projection function can serve as an activation function for NNs in L2O applications, since it not only constrains the NN’s output to lie within a convex set but also ensures that the output range exactly matches the feasible region—a critical requirement for learning the optimal solution. The primary challenge with employing the Hilbert projection function as an activation mechanism is that it is derivated from a convex program, which does not provide a closed-form expression for the gradient connecting the projected point to the raw NN output. To address this, we adopt a heuristic gradient derivation approach based on the Hadamard product and division. The main contributions of this paper are summarized as follows.
\begin{enumerate}
    \item We propose a Hilbert projection-based activation function that guarantees strict adherence to convex constraints for any neural network architecture, thereby ensuring that all outputs lie within the feasible region.
    \item Building upon the Hadamard product and division, we provide a heuristic gradient derivation method that explicitly computes the partial derivatives of the objective function with respect to the NN parameters. This enables the proposed L2O framework to handle arbitrary differentiable (non-)convex problems under convex constraints.
    \item Numerical simulations demonstrate that the proposed L2O approach outperforms traditional convex analysis-based methods and penalty-based learning schemes, highlighting its advantages in terms of both solution accuracy and computational efficiency.
\end{enumerate}

\section{Problem Formulation}
The problem under study involves optimizing a continuous differentiable (non-)convex objective function over a closed convex feasible region. Formally, the optimization problem is defined as follows.
\begin{problem}\label{problem-1}
\begin{align}
\min_{\bm{x}\in\mathcal{C}} \, f(\bm{x}),
\end{align}
\end{problem}
\noindent where $\mathcal{C} \subseteq \mathbb{C}^n$ represents the feasible region, which is a closed convex set, and $f(\bm{x}): \mathbb{C}^n \to \mathbb{C}$ is a continuous differentiable function that may be either convex or non-convex. The convexity of $\mathcal{C}$ ensures that any line segment connecting two points within the set lies entirely within the set, providing a well-defined and structured feasible region. This property is advantageous for constraint handling, as it simplifies the process of verifying feasibility and facilitates projection operations. However, the objective function $f(\bm{x})$ introduces complexity, particularly when it is non-convex. In non-convex optimization, the objective function can exhibit multiple local minima, saddle points, or other irregularities that make finding the global minimum a challenging task.

The inherent complexity of this problem arises from the interaction between the non-convexity of the objective function and the constraints defined by the closed convex set. When $f(\bm{x})$ is convex, the problem is computationally tractable under standard convex optimization frameworks, and the global minimum can be efficiently found using methods such as gradient descent or interior point techniques. However, when $f(\bm{x})$ is non-convex, the problem becomes significantly more challenging. The non-convexity of $f(\bm{x})$ implies that the optimization landscape may contain numerous local minima, making it difficult to distinguish between local and global optima. Standard gradient-based optimization methods are prone to getting trapped in local minima or saddle points, particularly when the initial point lies in a poorly conditioned region of the search space. Furthermore, this problem is classified as NP-hard when $f(\bm{x})$ is non-convex. NP-hardness implies that no polynomial-time algorithm is known to guarantee finding the global optimum for all instances of the problem. The complexity arises from the exponential growth of the search space with the problem dimension $n$, coupled with the difficulty of navigating the non-convex landscape. Even for relatively simple non-convex functions, identifying the global minimum may require evaluating an exponential number of candidate solutions, which becomes computationally infeasible for high-dimensional problems.

\section{Related Works}
Convex optimization problems have been extensively studied, resulting in a plethora of well-established algorithms capable of reliably finding global optimal solutions \cite{boyd2004convex}\cite{bertsekas2003convex}\cite{borwein2006convex}. Algorithms such as Newton's method, the steepest descent method, and the conjugate gradient method exemplify these advancements, leveraging the fundamental property that any local minimum of a convex optimization problem is also a global minimum. In contrast, solving non-convex optimization problems is inherently more complex due to the lack of this property. A widely adopted strategy involves transforming the non-convex problem into a sequence of approximated convex subproblems, which are iteratively solved. Representative techniques include the successive convex approximation algorithm \cite{hong2015decomposition4} and the majorization-minimization (MM) algorithm \cite{hunter2004tutorial5}. However, these methods can be computationally inefficient, particularly when appropriate initial points are unavailable. Moreover, designing suitable convex approximations for the original non-convex problem poses significant challenges.
The Difference of Convex Algorithm (DCA) \cite{horst1999dc6} is a widely utilized approach for solving non-convex optimization problems. DCA achieves linear convergence by representing the objective function as the difference of two convex functions, with the difference progressively diminishing over successive iterations. While DCA theoretically guarantees convergence to a local optimal solution, it has been observed to often converge to a global optimal solution in practice. Additionally, its computational efficiency makes it particularly suitable for large-scale optimization problems \cite{an2005dc7}.

Another prominent method, the Alternating Direction Method of Multipliers (ADMM) \cite{cai2017convergence8}\cite{nishihara2015general9}\cite{deng2016global10}, has gained significant attention for solving optimization problems with separable convex objectives. For non-convex problems, Bregman distance-based ADMM (Bregman ADMM) extends the traditional ADMM framework by incorporating the Bregman distance into the augmented Lagrangian function, thereby enhancing the subproblem-solving process \cite{wang2014convergence11}\cite{chen2016direct12}. Nevertheless, when bounded constraints are present, the direct application of ADMM continues to encounter significant challenges, limiting its effectiveness in such scenarios.In \cite{zhang2020proximal}, a gradient projection algorithm is introduced to address non-convex optimization problems on bounded polyhedra. This approach effectively resolves single-block optimization problems with bounded constraints, demonstrating improved performance over traditional methods.

Since the introduction of the neural dynamics algorithm by Hopfield et al. in 1985 \cite{hopfeild1985neural13}, neural network-based methods have been extensively applied to a variety of optimization problems. These methods leverage diverse approaches, including the Lagrange multiplier method, projection method, and penalty function method. For instance, a neural network algorithm utilizing the subgradient method was proposed in \cite{bian2009subgradient14} to address non-smooth and non-convex optimization problems. However, this approach requires the computation of an exact penalty factor, which significantly increases computational complexity. Similarly, a projection-based neural network algorithm was developed in \cite{gao2009new15}to solve optimization problems with variational inequality constraints, but its applicability is limited to problems with inequality constraints only.

Although the aforementioned methods have demonstrated progress in addressing non-convex optimization problems, the prediction errors inherent in neural networks can compromise the feasibility of the generated solutions and may even result in erroneous outputs. To mitigate constraint violations in predicted solutions, several studies have proposed enhancing penalty functions within the loss function \cite{cheng2019end16}\cite{pan2020deepopf17}\cite{zamzam2020learning18}\cite{kotary2021end19} or embedding the Karush-Kuhn-Tucker (KKT) conditions as equality constraints directly into the neural network model to improve solution performance \cite{nellikkath2021physics20}. However, due to the inherent prediction inaccuracies of neural networks, these approaches still fail to guarantee feasibility under the given constraints.
To address infeasible solutions, a gradient-based micro-correction method was proposed in \cite{donti2021dc321}. While this method provides some improvement, it cannot guarantee feasibility for general constraints.

In summary, existing approaches for ensuring the feasibility of neural network-based solutions either suffer from high computational complexity or fail to provide feasibility and optimality guarantees. Consequently, there remains a significant need for a general and computationally efficient method that ensures feasibility under general constraint conditions.

\section{Differentiable Projection-Based L2O Method}
\begin{figure*}
    \centering
    \includegraphics[width=0.95\linewidth]{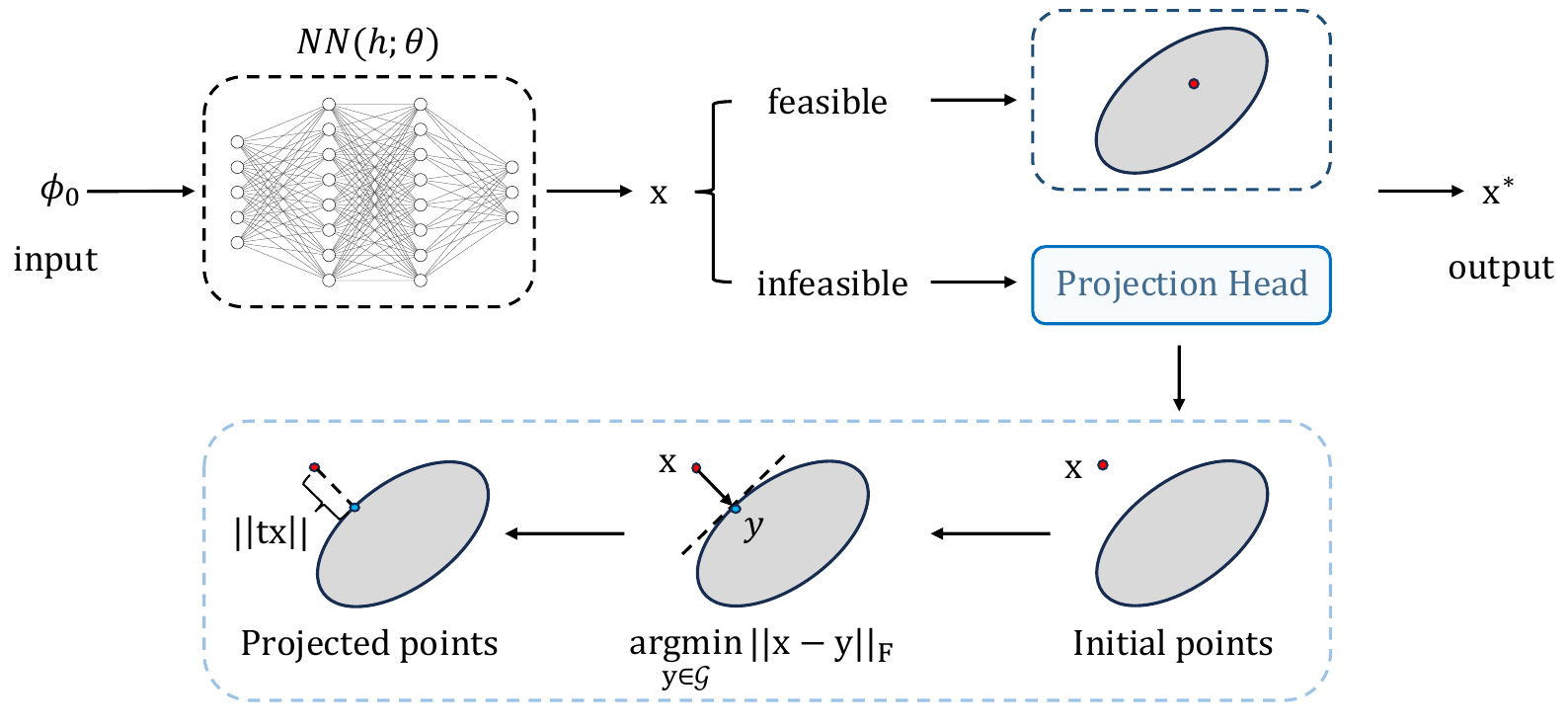}
    \caption{Illustration of projection-based L2O for convex constrained (non-)convex programming.}
    \label{fig-system}
\end{figure*}
\subsection{Convex Projection Theorem}
The projection theorem of closed convex sets is a fundamental result in convex analysis, providing a theoretical basis for projecting points onto convex sets, a widely used technique in optimization problems. This theorem asserts that for any point $\bm{x} \in \mathbb{C}^n$ outside a closed convex set $\mathcal{C} \subseteq \mathbb{C}^n$, there exists a unique point $\bm{y}^* \in \mathcal{C}$ that is closest to $\bm{x}$ in terms of Euclidean distance. Furthermore, the problem of finding this closest point can be formulated as a convex optimization problem, which is both computationally efficient and theoretically well-defined.

Mathematically, the projection of $\bm{x}$ onto $\mathcal{C}$ is the solution to the following optimization problem.
\begin{align}
\text{Proj}(\bm{y})=\bm{y}^* = \underset{\bm{y} \in \mathcal{C}}{\text{argmin}} \, \|\bm{y} - \bm{x}\|_2^2,
\end{align}
where $\|\cdot\|_2$ denotes the Euclidean norm. The objective function, $\|\bm{y} - \bm{x}\|_2^2$, is a strictly convex quadratic function of $\bm{y}$. 

\begin{lemma}\label{lemma-1}
    For a closed convex set $\mathcal{C}\in\mathbb{C}^n$, and $y\in \mathbb{C}^n$ the projection Proj$(\bm{y})$ is unique.
\end{lemma}

\begin{proof}
    To establish the uniqueness of the closest point, consider the properties of the Euclidean norm. Strict convexity implies that the objective function $\|\bm{y} - \bm{x}\|_2^2$ has a unique minimizer. Suppose, for contradiction, that there exist two distinct points $\bm{y}_1, \bm{y}_2 \in \mathcal{C}$ that both minimize the distance to $\bm{x}$. By the strict convexity of the Euclidean norm, any convex combination of these two points, $\bm{y}_\alpha = \alpha \bm{y}_1 + (1-\alpha) \bm{y}_2$ with $\alpha \in (0, 1)$, would yield a smaller distance to $\bm{x}$ than either $\bm{y}_1$ or $\bm{y}_2$, which contradicts the assumption that both points are minimizers. Therefore, the minimizer $\bm{y}^*$ must be unique. Moreover, the projection point $\bm{y}^*$ satisfies the orthogonality condition:
\begin{align}
(\bm{y}^* - \bm{x})^\top (\bm{z} - \bm{y}^*) \geq 0, \quad \forall \bm{z} \in \mathcal{C}.
\end{align}
This condition arises from the first-order optimality condition of the convex optimization problem and provides an important geometric interpretation: the vector from $\bm{x}$ to $\bm{y}^*$ is orthogonal to any vector pointing from $\bm{y}^*$ to another point in $\mathcal{C}$.
\end{proof}

The above analysis shows that we can use the projection theorem of convex sets. When NN outputs an infeasible solution, we can project it to the convex feasible region by solving a convex problem. Moreover, according to Lemma \ref{lemma-1}, the projection of an infeasible solution to a convex set is unique. Therefore, if we can make NN learn how the output of an infeasible solution affects the performance of solving a constrained problem, instead of simply discarding samples of infeasible solutions and keeping them out of training, we can improve the utilization efficiency of data and thus enhance the convergence speed of NN.

\subsection{Projection-Based Feasibility Guarantee Method}
\begin{algorithm}[h]
\caption{Projection-Based L2O Inference Method}
\begin{algorithmic}[1]
\REQUIRE Trained NN $\mathcal{H}$, coefficients $\bm{\phi}_0$ (objective) and $\bm{\phi}_i$ (constraints), feasible region $\mathcal{G}$
\ENSURE Feasible solution $\bm{x}$
\STATE \textbf{Input:} Coefficients $\bm{\phi}_0$ and $\bm{\phi}_i$ into the trained NN $\mathcal{H}$
\STATE Predict solution $\bm{x} = \mathcal{H}(\bm{\phi})$
\IF{$\bm{x} \in \mathcal{G}$}
    \STATE \textbf{Output:} $\bm{x}$ 
\ELSE
    \STATE Solve the projection problem:
    \begin{align*}
    \bm{y} = \underset{\bm{y} \in \mathcal{G}}{\text{argmin}} \|\bm{x} - \bm{y}\|_F
    \end{align*}
    \STATE \textbf{Output:} $\bm{y}$ 
\ENDIF
\end{algorithmic}\label{algorithm-1}
\end{algorithm}
To ensure the generality and broad applicability of the proposed approach, we deliberately avoided designing a NN architecture tailored to any specific problem. Instead, we adopted a MLP as the backbone network. This decision allows our method to generalize effectively across a wide range of optimization problems without being constrained by the structural limitations of specialized network architectures. Many wireless network optimization tasks, such as network resource allocation, edge computing management, and beamforming, share a common mathematical structure when solving non-convex optimization problems with convex feasible regions. These tasks primarily differ in the coefficients of the objective and constraint equations, which are determined by real-world parameters such as the number of resources, channel state information (CSI), and user requirements. To formalize this, the optimization problem is expressed as follows.
\begin{problem}\label{problem-2}
    \begin{align}
        &\min_{\bm{x}}\,\,f(\bm{x};\bm{\phi}_0),\label{obj-2}\\
        &\text{s.t.}\quad \bm{x}\in\mathcal{G},\tag{\ref{obj-2}a}\\
        &\qquad\;\, \mathcal{G}=\left\{\bm{x}\left|\begin{array}{l}
            g_i(\bm{x};\bm{\phi}_{i})\leq 0,\forall i\in\{1,\cdots,N\},\\
             h_i(\bm{x};\bm{\phi}_{i})= 0,\forall i\in\{N+1,\cdots,M\},
             \end{array}\right.\right\}\tag{\ref{obj-2}b}\\
        &\qquad\;\, \mathcal{G}\subseteq\mathcal{C},\tag{\ref{obj-2}c}
    \end{align}
\end{problem}
\noindent where $f(\cdot)$ represents a differentiable (non-)convex objective function, while $\bm{\phi}_0$ and $\bm{\phi}_i$ are problem-specific coefficients derived from wireless network parameters. The set $\mathcal{G}$ defines the feasible region of Problem \ref{problem-2}, and $\mathcal{C}$ denotes a larger convex set that contains $\mathcal{G}$. The convexity of $\mathcal{C}$ ensures computational efficiency during projection operations.

To simplify NN training and avoid the artificial general intelligence (AGI) assumption—where a single NN must handle multiple diverse optimization tasks—we propose training a dedicated NN for each specific task. Each NN focuses on a single network optimization problem, taking only relevant parameters such as CSI and user requirements as input, and outputs the optimal values of the optimization variables.To ensure the feasibility of the NN’s output, we employ a projection method whenever the NN outputs an infeasible solution $\bm{x}$. This infeasible solution is projected onto the nearest point within the feasible region by solving the following problem.
\begin{problem}\label{problem-3}
    \begin{align}
        &\min_{\bm{y}}\,\,\|\bm{x}-\bm{y}\|_{F},\label{obj-3}\\
        &\text{s.t.}\quad \bm{y}\in\mathcal{G},\tag{\ref{obj-3}a}\\
        &\qquad\;\, \mathcal{G}=\left\{\bm{y}\left|\begin{array}{l}
            g_i(\bm{y};\bm{\phi}_{i})\leq 0,\forall i\in\{1,\cdots,N\},\\
             h_i(\bm{y};\bm{\phi}_{i})= 0,\forall i\in\{N+1,\cdots,M\},
             \end{array}\right.\right\}\tag{\ref{obj-3}b}\\
        &\qquad\;\, \mathcal{G}\subseteq\mathcal{C},\tag{\ref{obj-3}c}
    \end{align}
\end{problem}
\noindent where $\|\bm{x} - \bm{y}\|_F$ represents the Frobenius norm, used to minimize the distance between the infeasible solution $\bm{x}$ and the feasible region. Since the feasible region $\mathcal{G}$ is a subset of the convex set $\mathcal{C}$, Problem \ref{problem-3} is a convex second-order programming problem. It can be efficiently solved using well-established methods such as Newton’s method, the KKT conditions, or interior point methods. By combining the task-specific NN design with the projection-based approach, our method ensures that solutions remain feasible while maintaining computational efficiency and adaptability to various wireless network optimization problems.

As is shown in Algorithm \ref{algorithm-1}, the projection-based approach for ensuring feasibility in NN outputs offers significant advantages over traditional methods, particularly in scenarios requiring efficient computation and real-time optimization. This section provides a detailed analysis of the key benefits of this method. One of the primary advantages of the projection method is its ability to ensure that all solutions output by the NN remain strictly within the feasible region throughout the entire optimization process, including during training. This guarantees strict adherence to the problem's constraints at all times, overcoming one of the fundamental limitations of traditional penalty-based methods. Penalty loss approaches approximate constraint enforcement by incorporating penalty terms into the loss function, which results in a probabilistic rather than absolute guarantee of constraint satisfaction. Constraint violations may still occur during or after training due to approximation errors, requiring significant computational effort for offline convergence to adequately penalize violations. In contrast, the projection-based approach directly corrects infeasible solutions online, ensuring immediate feasibility. This capability makes it particularly well-suited for online NN optimization, where fast and reliable decision-making is essential. Another notable benefit of the projection method lies in its computational efficiency. Solving the projection problem involves minimizing a simple quadratic objective function, $\|\bm{x} - \bm{y}\|_F^2$, which is computationally straightforward. This structure enables the use of established convex optimization techniques, including Newton's method, interior point methods, or even closed-form solutions in certain cases. When the constraints are linear or otherwise simple, the projection problem becomes a quadratic program (QP), which can be solved in polynomial time. This efficiency is particularly advantageous when compared to directly solving the original non-convex optimization problem, which often involves iterative algorithms prone to local minima and high computational costs. By reducing the complexity of feasibility enforcement, the projection method not only ensures constraint satisfaction but also minimizes the computational burden, making it suitable for large-scale, practical applications. The projection-based method is also highly versatile and broadly applicable. It requires only two fundamental conditions: the objective function must be differentiable, and the feasible region must be a convex set. These minimal requirements allow the method to be generalized to a wide range of optimization problems, including network resource allocation, edge computing management, and beamforming in wireless networks. Its universality makes it a valuable tool in diverse optimization and machine learning tasks, particularly in domains where strict and efficient constraint enforcement is crucial.

\subsection{Unsupervised Training Method}
\begin{algorithm}[h]
\caption{Projection-Based L2O Training Method}
\begin{algorithmic}[1]
\REQUIRE Objective function $f(\bm{x}; \bm{\phi}_0)$, constraints $g_i(\bm{x}; \bm{\phi}_i) \leq 0$, $h_i(\bm{x}; \bm{\phi}_i) = 0$, initial NN parameters $\bm{\theta}_0$, learning rate $\alpha$, gradient clipping threshold $\beta$, perturbation $\bm{\epsilon}$
\ENSURE Trained network parameters $\bm{\theta}$
\STATE Initialize $\bm{\theta}_0$
\FOR{$k = 1, 2, \dots, K$ (number of training iterations)}
    \STATE Sample coefficients $\bm{\phi}_0$ and $\bm{\phi}_i$ from the training dataset
    \STATE Predict solution $\bm{x} = \mathcal{H}(\bm{\phi}; \bm{\theta}_{k-1})$
    \IF{$\bm{x} \in \mathcal{G}$}
        \STATE Compute the gradient:
        \begin{align*}
        \bm{p}_k = \nabla_{\bm{x}} f(\bm{x}) \cdot \nabla_{\bm{\theta}} \bm{x}
        \end{align*}
    \ELSE 
        \STATE Solve the projection problem:
        \begin{align*}
        \bm{y} = \underset{\bm{y} \in \mathcal{G}}{\text{argmin}} \|\bm{x} - \bm{y}\|_F
        \end{align*}
        \STATE Compute the scaling factor:
        \begin{align*}
        \bm{t} = \bm{y} \oslash (\bm{x} + \bm{\epsilon})
        \end{align*}
        \STATE Compute the gradient:
        \begin{align*}
        \bm{p}_k = \nabla_{\bm{t} \circ \bm{x}} f(\bm{t} \circ \bm{x}) \cdot \nabla_{\bm{\theta}} \bm{x}
        \end{align*}
    \ENDIF
    \STATE Clip the gradient:
    \begin{align*}
    \bm{p}_k = Clip(\bm{p}_k, \beta)
    \end{align*}
    \STATE Update the NN parameters:
    \begin{align*}
    \bm{\theta}_k = \bm{\theta}_{k-1} + \alpha\; \text{optim}(\bm{p}_k,\cdots,\bm{p}_0)
    \end{align*}
\ENDFOR
\STATE \textbf{Output:} Trained network parameters $\bm{\theta}_K$
\end{algorithmic}
\end{algorithm}
To overcome the challenges of NN training in the L2O method relying on the optimal solution as a label, and the slow convergence speed of traditional unlabeled training methods such as reinforcement learning, we adopted an unsupervised training method. That is, by using the chain rule, we explicitly give the gradient of the optimization objective function with respect to the NN parameters. When the solution output by NN is feasible, we can update the NN parameters through the following formula.
\begin{align}
    &\bm{\theta}_{k} = \bm{\theta}_{k-1}+\alpha\; \text{optim}\left(\bm{p}_{k},\cdots,\bm{p}_{0}\right),\\
    &\bm{p}_{k} = \nabla_{\bm{x}}f(\bm{x})|_{\bm{x}=\mathcal{H}(\bm{\phi};\bm{\theta})}\nabla_{\bm{\theta}}\bm{x},\label{gradient-feasible}
\end{align}
where optim($\cdot$) is the optimizer used to update the parameter $\bm{\theta}$ of the NN $\mathcal{H}(\cdot)$, which can be the stochastic gradient descent if optim$\left(\bm{p}_{k},\cdots,\bm{p}_{0}\right)=\bm{p}_k$, moment, and Adam methods. The update direction $\bm{p}$ can be obtained according to \eqref{gradient-feasible} since the $f(\cdot)$ is differentiable, and the second term can be calculated through the gradient backward propagation of the NN.

When the initial output of the NN is infeasible, directly applying the gradient calculation based on \eqref{gradient-feasible} is not possible. This limitation arises because the objective function $f(\cdot)$ is evaluated using the projected variables $\bm{y}$, which depend on $\bm{x}$ through the solution to Problem \ref{problem-3}. Consequently, the gradient of $f(\cdot)$ with respect to $\bm{x}$ must be computed using the chain rule. However, this introduces a significant challenge, as there is often no explicit analytical expression that directly links $\bm{y}$ to $\bm{x}$. The complexity of the constraints in the optimization problem frequently necessitates iterative methods to obtain $\bm{y}$, making it impractical to compute the gradient $\nabla_{\bm{x}} \bm{y}$ directly. In cases involving non-linear or complex constraints, solving the Karush-Kuhn-Tucker (KKT) conditions analytically is infeasible, further complicating the derivation of $\nabla_{\bm{x}} f$.

To overcome these difficulties, we adopt a heuristic approach to approximate the relationship between $\bm{y}$ and $\bm{x}$. Given that $\bm{y}$ and $\bm{x}$ have the same dimensions, we calculate an element-wise scaling factor $\bm{t}$ by dividing $\bm{y}$ by $\bm{x}$, element by element. To prevent division by zero, a small perturbation $\bm{\epsilon}$ is added to $\bm{x}$ during the computation, ensuring numerical stability. This calculation is expressed as:
\begin{align}
\bm{t} = \bm{y} \oslash (\bm{x} + \bm{\epsilon}),
\end{align}
where $\oslash$ represents the Hadamard (element-wise) division. The scaling factor $\bm{t}$ serves as an approximation of the influence of $\bm{x}$ on the projected variables $\bm{y}$, enabling us to approximate the gradient of $f(\cdot)$ effectively.

The gradient of $f(\cdot)$ with respect to $\bm{x}$ is then computed by applying the chain rule to account for the dependency of the objective function on both $\bm{x}$ and $\bm{t}$. Specifically, the gradient $\nabla_{\bm{t} \circ \bm{x}} f(\bm{t} \circ \bm{x})$ is expanded as:
\begin{align}
\nabla_{\bm{t} \circ \bm{x}} f(\bm{t} \circ \bm{x}) = \frac{\partial f}{\partial (\bm{t} \circ \bm{x})},
\end{align}
where $\bm{t} \circ \bm{x}$ represents the Hadamard (element-wise) product. For each dimension $i$, the gradient is computed as:
\begin{align}
\frac{\partial f}{\partial (\bm{t} \circ \bm{x})_i} = \frac{\partial f}{\partial z_i}, \quad \text{where } z_i = t_i x_i.
\end{align}
By further applying the chain rule, the dependency of $f$ on both $\bm{t}$ and $\bm{x}$ is incorporated:
\begin{align}
\nabla_{\bm{x}} f(\bm{t} \circ \bm{x}) = \nabla_{\bm{z}} f(\bm{z}) \cdot \frac{\partial \bm{z}}{\partial \bm{x}},
\end{align}
where $\bm{z} = \bm{t} \circ \bm{x}$, and $\frac{\partial \bm{z}}{\partial \bm{x}}$ is calculated numerically.

To ensure numerical stability during parameter updates, we introduce gradient clipping. The parameters $\bm{\theta}$ of the NN are updated iteratively as:
\begin{align}
\bm{\theta}_k = \bm{\theta}_{k-1} + \alpha \; \text{optim}\left(Clip(\bm{p}_k), \cdots, Clip(\bm{p}_0)\right),
\end{align}
where $\bm{p}_k$ represents the calculated gradient:
\begin{align}
\bm{p}_k = \nabla_{\bm{t} \circ \bm{x}} f(\bm{t} \circ \bm{x}) \big|_{\bm{x} = \mathcal{H}(\bm{\phi}; \bm{\theta})} \nabla_{\bm{\theta}} \bm{x}.
\end{align}
The function $Clip(\cdot)$ ensures that the gradient values remain within a defined range $[- \beta, \beta]$. This is essential to prevent gradient explosion, which can occur when elements of $\bm{x}$ approach zero, leading to excessively large gradients. The clipping operation ensures that the NN remains stable during training, facilitating efficient convergence. This heuristic approach, combining the use of element-wise scaling, chain rule approximations, and gradient clipping, provides an effective means of handling infeasible initial outputs. By maintaining numerical stability and leveraging scalable computations, the method ensures robust and efficient NN training even in the presence of complex constraints.

The proposed training method enables the NN to learn how both feasible and infeasible outputs impact the performance of the solution, thereby enhancing its ability to optimize under complex constraints. For feasible outputs, the gradient calculation directly reflects the objective function, allowing the NN to iteratively refine its parameters to map input features, such as wireless network parameters, to optimal solutions within the feasible region. For infeasible outputs, the projection-based approach ensures that all solutions lie within the feasible region by leveraging the projected variables $\bm{y}$. The heuristic scaling factor $\bm{t}$ encodes the relationship between the original outputs $\bm{x}$ and their projections $\bm{y}$, enabling the NN to approximate the performance impact of corrections for infeasibility. This approach, combined with the chain rule for gradient propagation and gradient clipping to maintain stability, ensures that the NN effectively internalizes the effects of constraint violations while optimizing its output. By learning from both feasible and infeasible cases, the NN develops a comprehensive understanding of the optimization landscape, improving its generalization and ensuring high-quality, constraint-satisfying solutions in complex applications such as wireless network optimization.

\section{Applications on Beamforming Optimization}
\subsection{Security Communication Rate Under Quality of Service Constraints}
We consider a downlink communication system in which $N$ senders each serve one legitimate user and one eve user. Let $\bm{h}_{i,i} \in \mathbb{C}^{K \times 1}$ represent the channel response vector from sender $i$ (equipped with $K$ antennas) to its corresponding legitimate user $i$, who is assumed to have a single antenna. Similarly, let $\bm{g}_{i,k} \in \mathbb{C}^{K \times 1}$ denote the channel response between sender $i$ and the eve user $k$. All senders operate over the same frequency band, creating co-channel interference among simultaneous transmissions. To quantify the received signal quality, we define the signal-to-interference-plus-noise ratio (SINR). For legitimate user $i$, the SINR is given by
\begin{align}
\mathrm{SINR}_{i} \;=\; \frac{\bigl|\bm{w}_{i}^{H}\bm{h}_{i,i}\bigr|^{2}}{\displaystyle \sum_{j \neq i}\bigl|\bm{w}_{j}^{H}\bm{h}_{j,i}\bigr|^{2} \;+\; \sigma^{2}},
\end{align}
where $\bm{w}_{i} \in \mathbb{C}^{K \times 1}$ is the beamforming vector used by sender $i$, and $\sigma^{2}$ denotes the noise power. For the eve user $k$, the SINR is modeled as
\begin{align}
\mathrm{SINR}_{k} \;=\; \frac{\bigl|\bm{w}_{k}^{H}\bm{g}_{i,k}\bigr|^{2}}{\displaystyle \sum_{j \neq i}\bigl|\bm{w}_{j}^{H}\bm{g}_{j,k}\bigr|^{2} \;+\; \sigma^{2}},
\end{align}
reflecting the possibility of eavesdropping on signals originating from sender $i$. To maintain a target quality of service (QoS) for legitimate users, the system must satisfy $\mathrm{SINR}_{i} \geq \gamma$, while reducing the likelihood of successful eavesdropping by minimizing the eve users’ SINR. These objectives can be cast into the following optimization problem.
\begin{problem}\label{problem-4}
    \begin{align}
        &\min_{\bm{w}} \sum_{k=1}^{N} \frac{\bigl|\bm{w}_{k}^{H}\bm{g}_{i,k}\bigr|^{2}}{\displaystyle \sum_{j \neq i}\bigl|\bm{w}_{j}^{H}\bm{g}_{j,k}\bigr|^{2} \;+\; \sigma^{2}},\label{obj-task-1}\\
        &\text{s.t.} \quad \frac{\bigl|\bm{w}_{i}^{H}\bm{h}_{i,i}\bigr|^{2}}{\displaystyle \sum_{j \neq i}\bigl|\bm{w}_{j}^{H}\bm{h}_{j,i}\bigr|^{2} \;+\; \sigma^{2}}\;\geq\;\gamma_i,\quad \forall i\in \{1,\cdots, N\}.\tag{\ref{obj-task-1}a}
    \end{align}
\end{problem}
\noindent where the goal is to find the set of beamforming vectors $\{\bm{w}_{i}\}$ that maximize the SINR of legitimate users while suppressing the SINR of eve users. Ensuring $\mathrm{SINR}_{i} \ge \gamma$ for each legitimate user $i$ introduces constraints that can be converted into a second-order cone (SOC) form, which is convex. However, minimizing the fractional structure of $\mathrm{SINR}_{k}$ for eve users results in a non-convex objective. By rewriting $\mathrm{SINR}_i \geq \gamma$ in a square root form, the inequality becomes suitable for second-order cone programming (SOCP), thereby preserving a convex constraint space. Consequently, the overall problem poses a non-convex optimization challenge because of the objective term that aims to reduce eavesdropping efficiency. To solve Problem \ref{problem-4} using an L2O framework, all channel responses $\bm{h}$ and $\bm{g}$ are treated as part of the parameter $\bm{\phi}$, which encapsulates legitimate and eve channel conditions. These parameters are fed into a trained NN, whose outputs are the beamforming vectors $\{\bm{w}_{i}\}$. If the NN’s output does not satisfy the required QoS constraints or fails to meet the SOC conditions, it is declared infeasible, prompting a projection step that maps the beamforming vector onto the closest point in the feasible region. This ensures that the final solution respects the system’s constraints, while the NN refines its beamforming strategy in response to evolving channel conditions.

\subsection{Green Communication Optimization Under QoS Constraints}
We consider a downlink communication system comprising a single sender and $N$ receiving users, each associated with a legitimate communication link. The channel response between the sender $i$ and the legitimate user $j$ is denoted by $\bm{h}_{i,j} \in \mathbb{C}^{K \times 1}$, where $K$ represents the number of antennas equipped by the sender. Each receiving user is equipped with a single antenna. In this model, all senders operate over the same frequency band, resulting in potential co-channel interference among simultaneous transmissions.

To evaluate the quality of communication for each user, we define the signal-to-interference-plus-noise ratio (SINR). For any legitimate user $i$, the SINR is given by
\[
\mathrm{SINR}_i = \frac{\left| \bm{w}_i^H \bm{h}_{i,i} \right|^2}{\sum_{j \neq i} \left| \bm{w}_j^H \bm{h}_{j,i} \right|^2 + \sigma^2},
\]
where $\bm{w}_i \in \mathbb{C}^{K \times 1}$ denotes the beamforming vector employed by sender $i$, and $\sigma^2$ represents the noise power at the receiving end. The primary objective is to ensure that the SINR of legitimate users meets or exceeds a predefined threshold $\gamma$, thereby guaranteeing the desired QoS.

In addition to maintaining QoS for legitimate users, it is imperative to minimize the power consumption associated with signal transmission. This objective is formalized by minimizing the total transmission power of the senders, expressed as the sum of the squared magnitudes of the beamforming vectors.
\[
\min_{\bm{w}} \sum_{i=1}^{N} \left\| \bm{w}_i \right\|_2^2.
\]
However, reducing the number of active antennas is equally important to lower the inherent energy consumption of operating multiple antennas. To achieve this, we introduce a sparsity constraint on the beamforming vectors $\bm{w}_i$. The sparsity is enforced by incorporating an $L_p$-norm regularization term into the optimization problem, where $p$ is chosen to promote sparsity. Specifically, we employ the $L_{0.5}$-norm as the regularization term, which effectively encourages sparsity in the beamforming vectors:
\[
\min_{\bm{w}} \sum_{i=1}^{N} \left\| \bm{w}_i \right\|_2^2 + \alpha \left\| \bm{w}_i \right\|_{0.5},
\]
where $\alpha$ is a regularization parameter that balances the trade-off between minimizing transmission power and promoting sparsity. The $L_{0.5}$-norm is chosen for its superior sparsity-inducing properties compared to the $L_1$-norm, despite being non-convex for $p < 1$. Traditional optimization methods struggle with such non-convex norms; however, the L2O framework is well-suited to handle non-differentiable and non-convex objectives, making it an ideal approach for this problem.

The optimization problem can be formally stated as follows:
\begin{problem}\label{problem-5}
    \begin{align}
        &\min_{\bm{w}} \sum_{i=1}^{N} \left\| \bm{w}_i \right\|_2^2 + \alpha \left\| \bm{w}_i \right\|_{0.5},\\
        &\text{s.t.} \quad \frac{\left| \bm{w}_i^H \bm{h}_{i,i} \right|^2}{\sum_{j \neq i} \left| \bm{w}_j^H \bm{h}_{j,i} \right|^2 + \sigma^2} \geq \gamma_i, \quad \forall i \in \{1, \cdots, N\}.
    \end{align}
\end{problem}
This formulation encapsulates a non-convex objective aimed at minimizing the SINR of eve users, thereby reducing the likelihood of successful eavesdropping. Concurrently, the constraints ensure that the SINR for each legitimate user $i$ remains above the threshold $\gamma_i$, thereby satisfying the QoS requirements. The non-convexity of the objective function arises from the fractional SINR terms, which are challenging to optimize using traditional methods. Nevertheless, the constraints can be reformulated into SOC representations as the above section, rendering them convex and thus computationally tractable within a convex optimization framework.

To address the optimization problem outlined in Problem \ref{problem-5} using the L2O method, all channel responses $\bm{h}_{i,j}$ and $\bm{g}_{i,k}$ are encapsulated within the parameter vector $\bm{\phi}$, which represents both legitimate and eve channel conditions. These channel parameters serve as inputs to a trained NN, which outputs the beamforming vectors $\{\bm{w}_i\}$. If the NN's output fails to satisfy the QoS constraints or violates the SOC conditions, the solution is deemed infeasible. In such cases, a projection problem is solved to map the infeasible beamforming vectors onto the nearest feasible region. This projection ensures that the final beamforming vectors adhere to all system constraints, thereby guaranteeing that legitimate users' SINR requirements are met while minimizing interference to eve users. As the system operates under varying channel conditions, the NN continuously refines its beamforming strategy through the L2O framework, enhancing its ability to generate high-quality, constraint-satisfying solutions efficiently.

\section{Numerical Results}
\subsection{Simulation Setting}
In this section, we experimentally validate the effectiveness of the proposed algorithm by generating 10,000 sets of randomly sampled channel impulse responses, where both $\bm{h}$ and $\bm{g}$ follow circularly symmetric complex Gaussian distributions $\mathcal{CN}(\bm{0}, \bm{I})$. The NN employed is a four-layer MLP with ReLU activation in the hidden layers and no activation in the final layer. During training, we adopt the Adam optimizer with a learning rate of 0.01 and a batch size of 64. In the projection step, we utilize cvxpy \cite{grant2014cvx}  to solve Problem \ref{problem-3}, thereby finding the nearest feasible point whenever the NN output is infeasible. This solution is then taken as the NN’s final output.

To benchmark the performance of our proposed approach, we compare it against three methods.
\begin{itemize}
    \item \textbf{Penalty Loss}: This method adds a penalty term in the NN loss function for constraint violations, aiming to guide the NN toward feasible solutions.
    \item Success Convex Approximation (\textbf{SCA}): This approach replaces non-convex terms with convex approximations, iteratively refining the solution to approximate feasibility and optimality.
    \item Trust Region Constraint (\textbf{Trust-Constr}): This algorithm updates solutions within a bounded trust region around the current point, ensuring incremental feasibility and stability in each iteration.
\end{itemize}

\subsection{Eve Users' SINR Minimization}
\begin{figure}[h]
    \centering
    \includegraphics[width=1.0\linewidth]{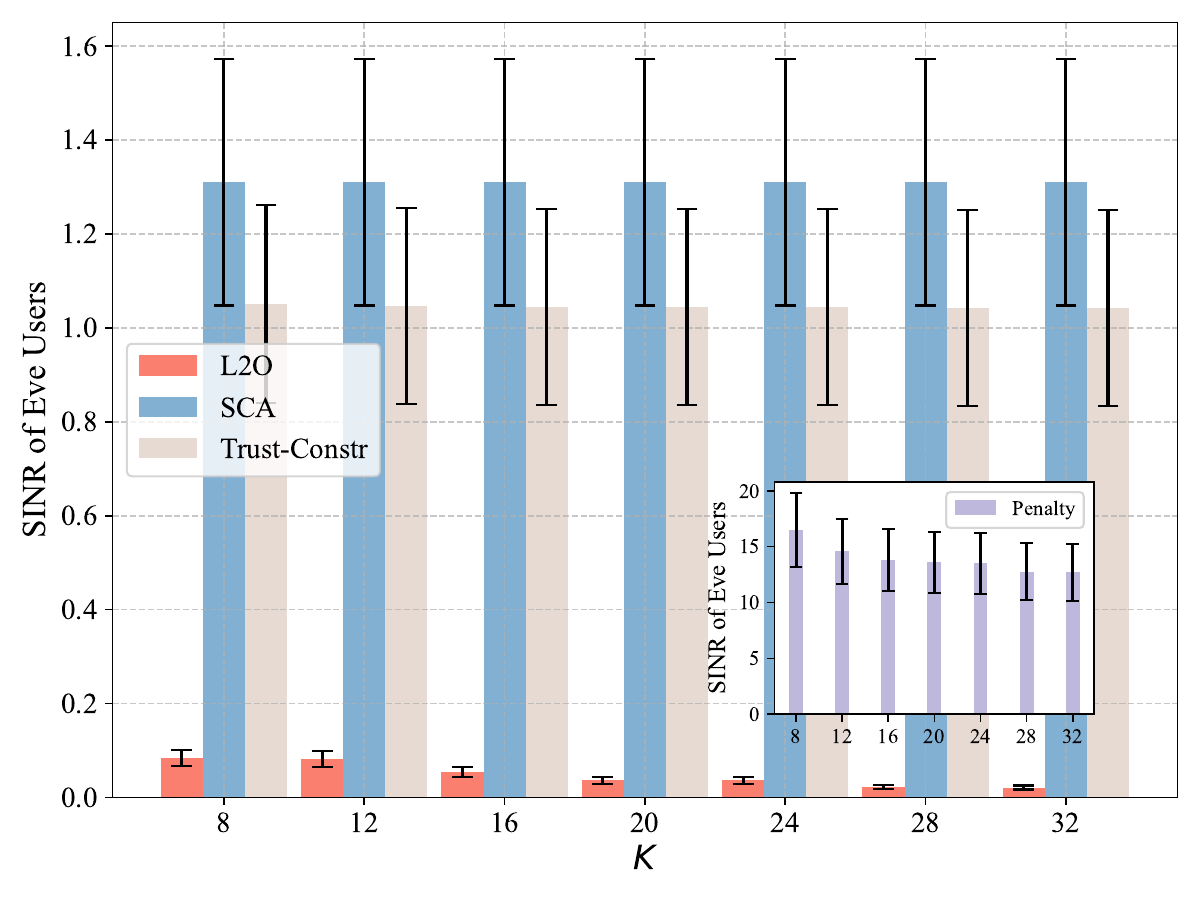}
    \caption{The comparison of different algorithms on $K$.}
    \label{fig-pls-antenna}
\end{figure}
\begin{figure}[h]
    \centering
    \includegraphics[width=1.0\linewidth]{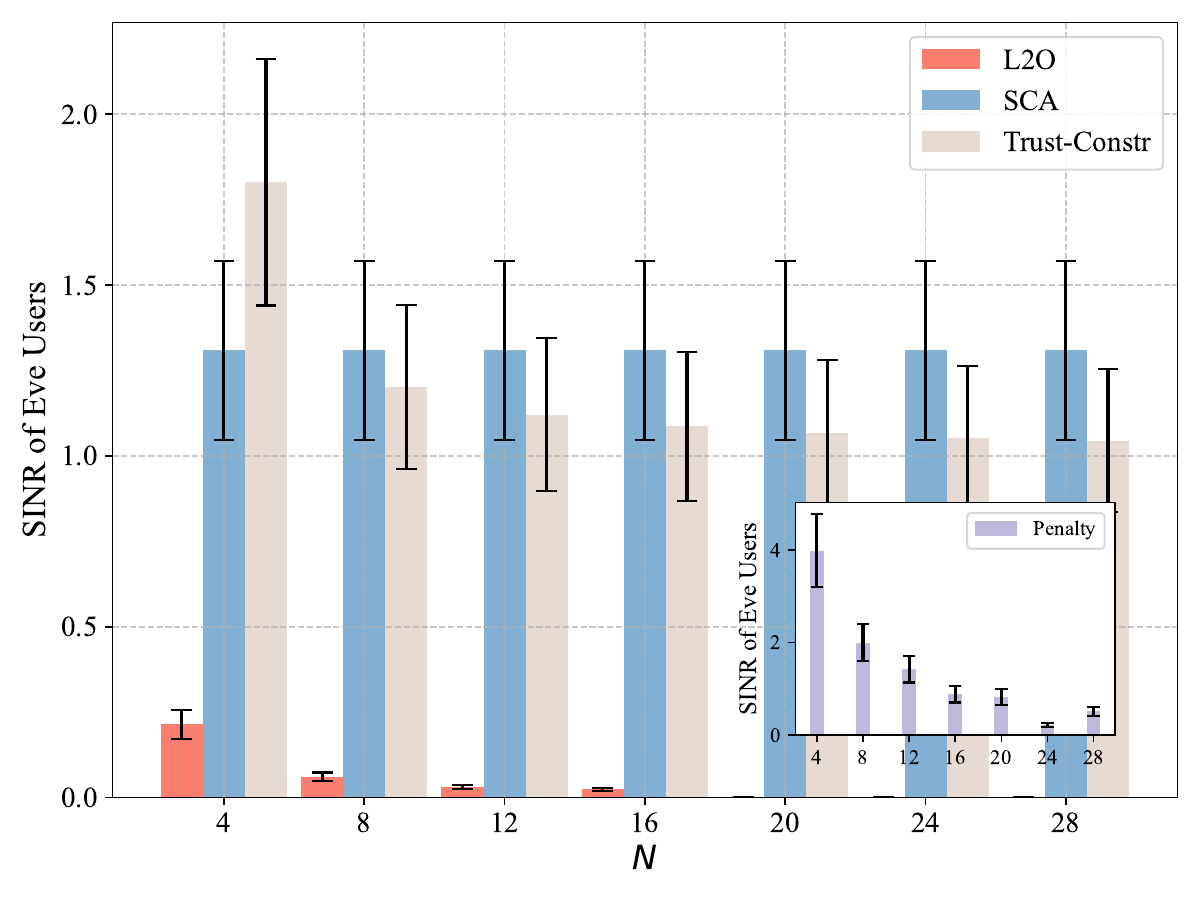}
    \caption{The comparison of different algorithms on $N$.}
    \label{fig-pls-users}
\end{figure}

\begin{figure}[h]
    \centering
    \includegraphics[width=1.0\linewidth]{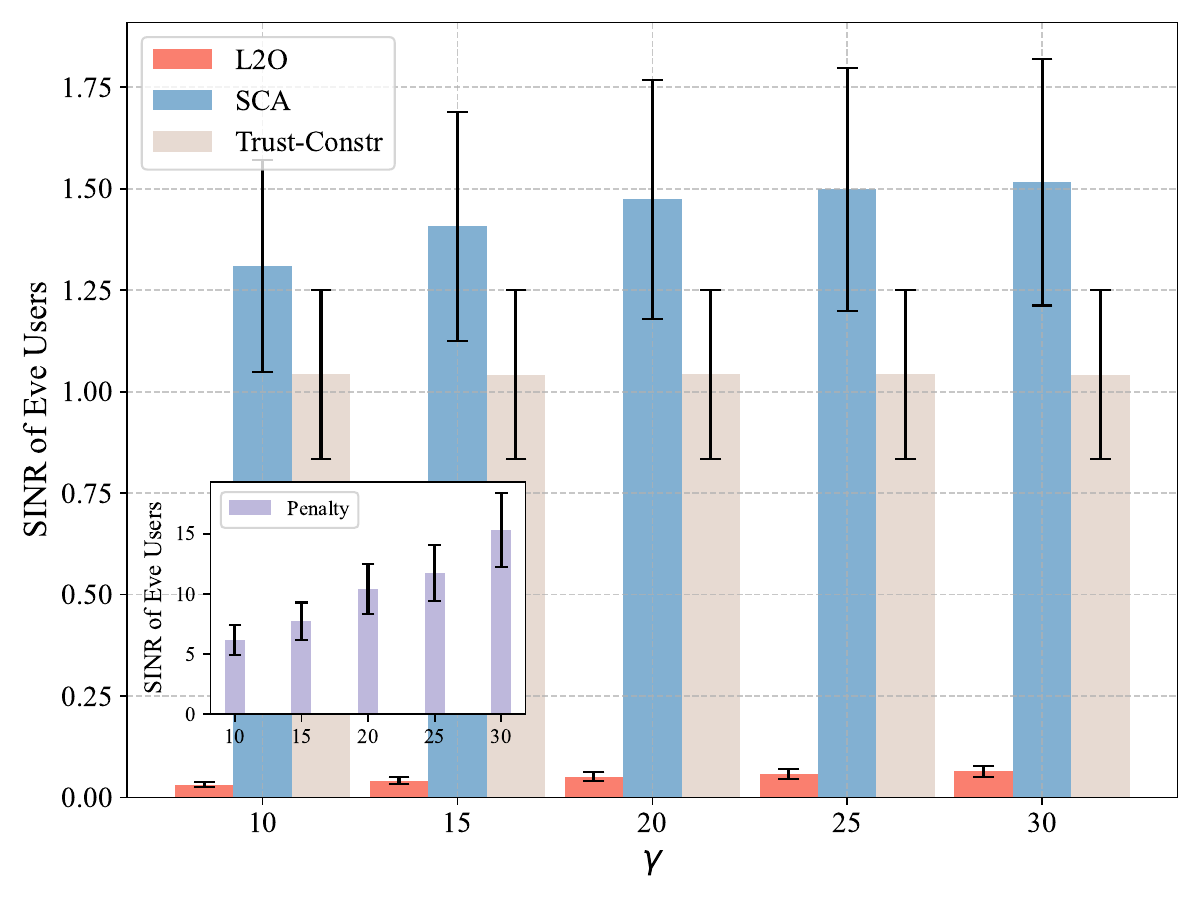}
    \caption{The comparison of different algorithms on $\gamma$.}
    \label{fig-pls-gamma}
\end{figure}
In this subsection, we analyze the performance of the proposed projection-based L2O method in comparison with several baseline algorithms under the default settings of $K = 16$, $N = 8$, and $\gamma = 10$. Fig.~\ref{fig-pls-antenna} illustrates the relationship between the number of antennas ($K$) and the eavesdropping SINR of an eve user when optimizing the beamforming vector using different algorithms. As $K$ increases, the SINR of the eve user decreases across all algorithms, a trend that can be attributed to the improved directivity of the transmitted beams. With more antennas, the sender is able to focus its transmission more effectively, making it more challenging for the eve user to intercept the signal. Notably, the proposed L2O method consistently achieves the lowest SINR for the eve user compared to baseline methods such as SCA and the trust region method, highlighting its superior performance in mitigating eavesdropping.

The results further underscore the effectiveness of the proposed projection method and its unsupervised training strategy. By allowing the NN to learn the impact of its output on constraint optimization, the L2O framework ensures constraint feasibility while leveraging the NN’s ability to extract critical features from channel impulse responses. This dual capability results in significant SINR reductions for the eve user, reinforcing the robustness of the L2O approach. As the number of antennas increases, the SINR reduction achieved by the L2O method outpaces that of the comparison algorithms, demonstrating its ability to exploit the characteristics of large-scale antenna systems effectively. This feature makes the proposed method particularly well-suited for scenarios involving massive multiple-input multiple-output (MIMO) systems.

An additional advantage of the proposed L2O method is its exceptional stability. Under identical conditions of antenna scale, user number, and other hyperparameters, the L2O method exhibits minimal performance variance across different channel realizations of $\bm{h}$ and $\bm{g}$. This stability is further corroborated by Figs.~\ref{fig-pls-users} and \ref{fig-pls-gamma}, where the L2O method demonstrates consistently low variance in performance across varying values of $N$ and $\gamma$. Such stability is crucial in practical deployments, where robustness and reliability are paramount.

Fig.~\ref{fig-pls-users} reveals a noteworthy phenomenon: as the number of users $N$ increases, the optimized beamforming vector significantly reduces the SINR of the eve user. This observation aligns with the concept of symbiotic security in physical layer security, where an increase in the number of users allows the sender to leverage beam cooperation to suppress the SINR of the eve user further. By simultaneously maintaining the QoS of legitimate users, the sender enhances information security, showcasing the potential of user cooperation in improving multi-user network security.

Finally, Fig.~\ref{fig-pls-gamma} examines the impact of varying the QoS threshold $\gamma$ on the SINR of the eve user. As $\gamma$ increases, the SINR of the eve user also rises gradually. This can be explained by the fact that, for a fixed number of antennas and users, a higher $\gamma$ forces the sender to transmit with greater power to meet the QoS requirements of legitimate users. Consequently, this increased transmission power inadvertently raises the SINR at the eve user’s location. These results underscore the trade-offs between QoS requirements and security performance, further emphasizing the adaptability of the L2O method in achieving optimal performance across varying network conditions.

\subsection{Transmission Power Minimization}
\begin{figure}[h]
    \centering
    \includegraphics[width=1.0\linewidth]{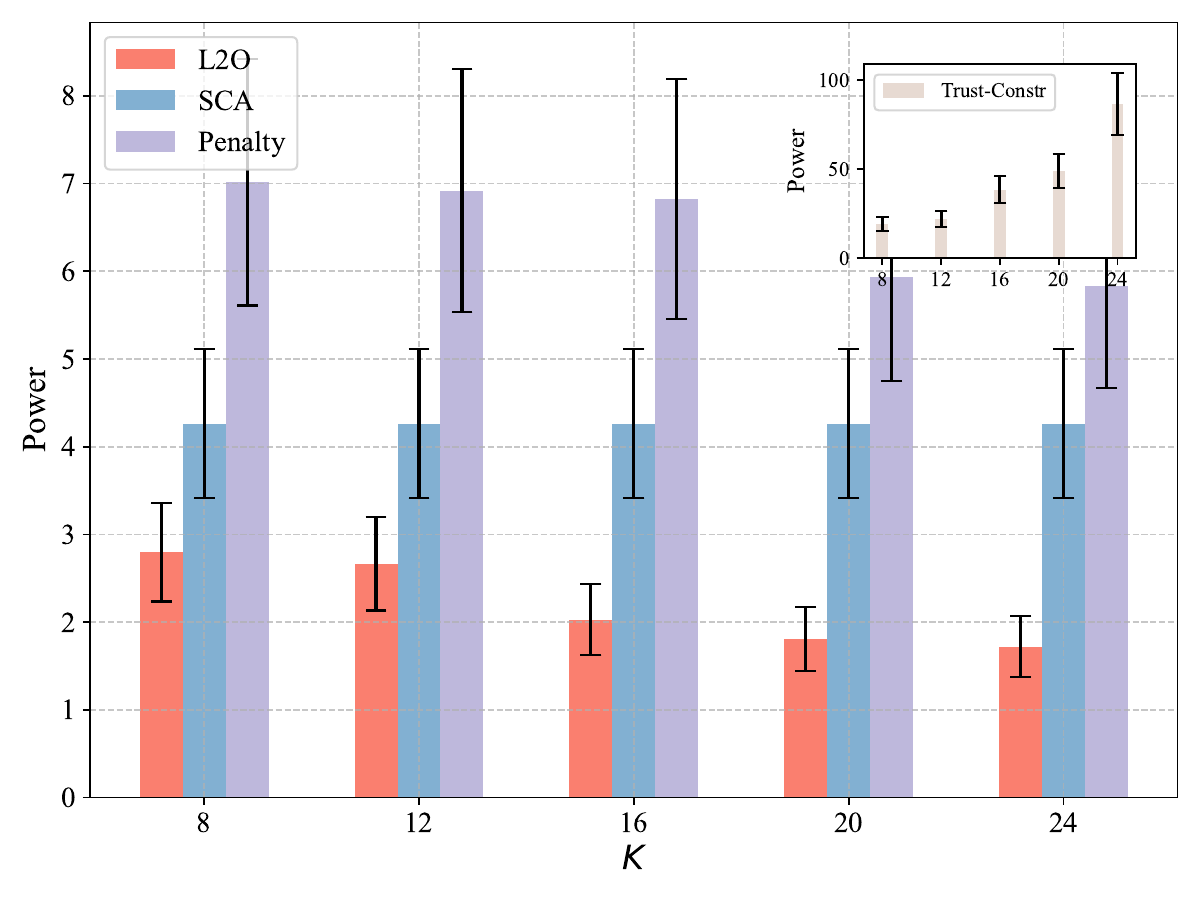}
    \caption{The power cost comparison of different algorithms on $K$.}
    \label{fig-power-antenna}
\end{figure}
\begin{figure}[h]
    \centering
    \includegraphics[width=1.0\linewidth]{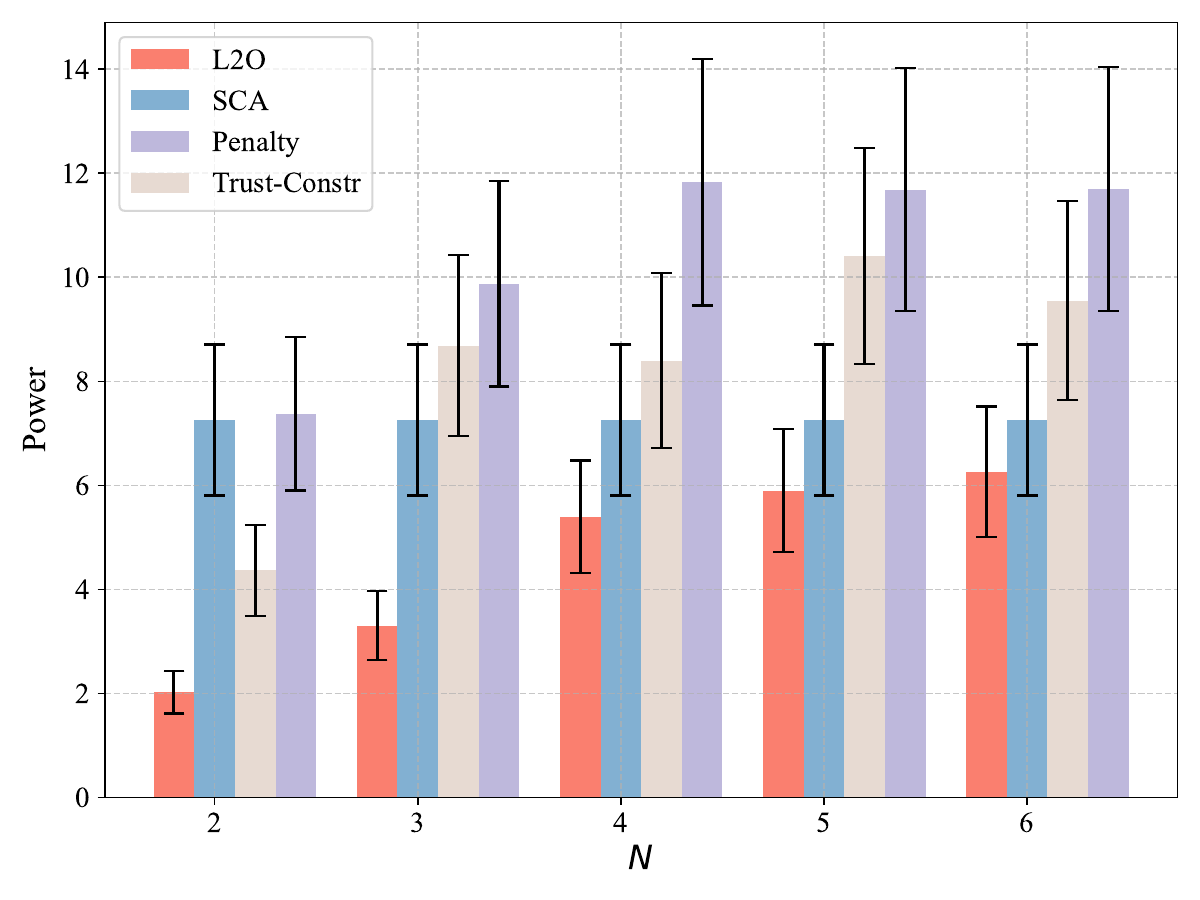}
    \caption{The power cost comparison of different algorithms on $N$.}
    \label{fig-power-users}
\end{figure}

\begin{figure}[h]
    \centering
    \includegraphics[width=1.0\linewidth]{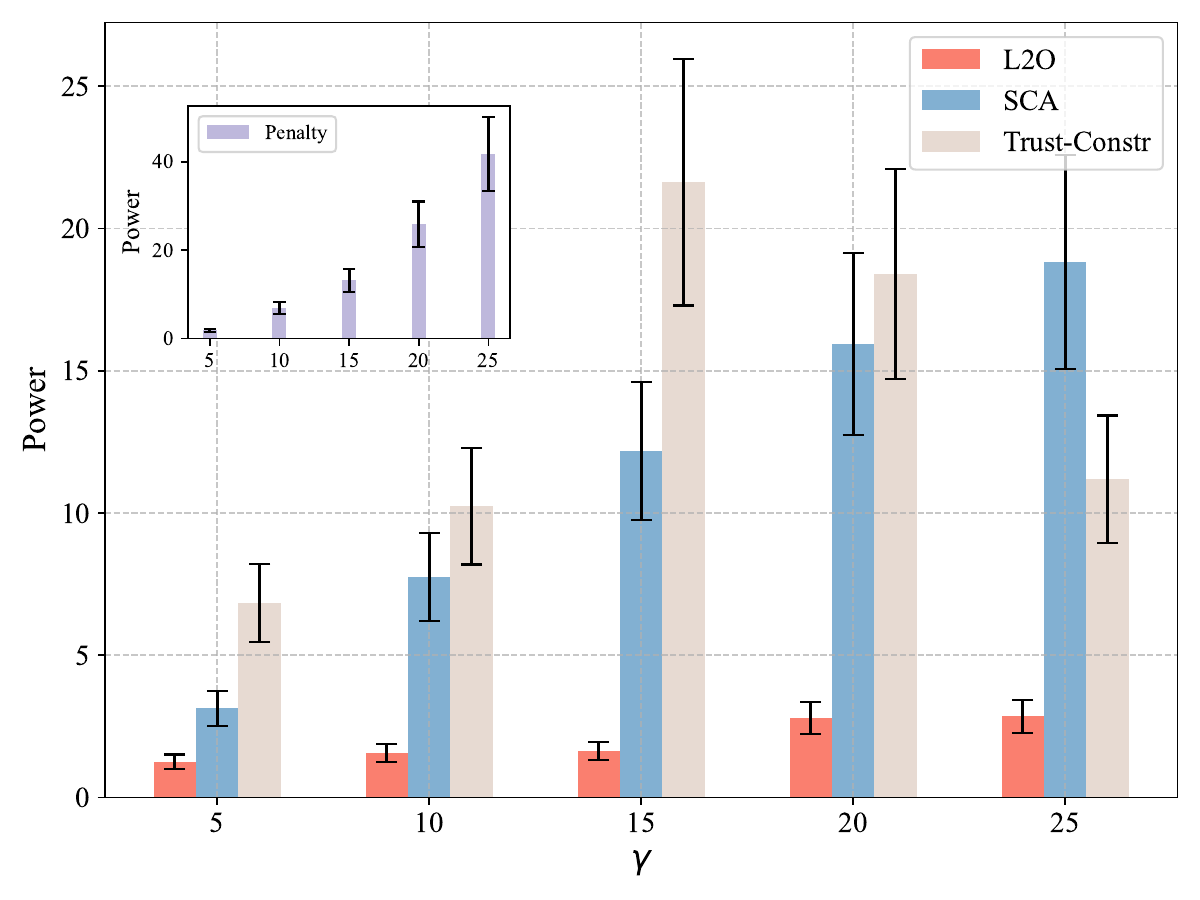}
    \caption{The power cost comparison of different algorithms on $\gamma$.}
    \label{fig-power-gamma}
\end{figure}

In this subsection, we analyze the performance of power consumption under SINR constraints for the proposed projection-based L2O method in comparison with several baseline algorithms under the default settings of $K = 16$, $N = 8$, and $\gamma = 10$. Fig.~\ref{fig-power-antenna} illustrates that, as the number of antennas increases, the energy consumption of all algorithms, except Trust-Constr, gradually decreases. This trend can be attributed to the increased directivity of the beam emitted by the sender with more antennas, which reduces beam interference. Consequently, less transmission power is required to meet user service demands. In contrast, Trust-Constr exhibits the opposite trend, highlighting the limitations of traditional non-convex optimization algorithms based on convex analysis. These limitations hinder the efficient convergence of such algorithms when solving diverse non-convex optimization problems. It is noteworthy that the projection-based L2O method proposed in this paper consumes the least energy across all antenna configurations. Additionally, this method demonstrates the most significant reduction in average energy consumption as the number of antennas increases. This indicates that the proposed algorithm is more adept at learning the characteristics of constrained non-convex optimization problems, allowing for better beam optimization. Moreover, Fig.~\ref{fig-power-antenna} shows that the variance in energy consumption for the projection-based L2O method decreases as the number of antennas increases, further supporting its robustness and efficiency in handling varying antenna configurations.

Fig.~\ref{fig-power-users} illustrates the variation in energy consumption as the number of users increases. It is evident that the projection-based L2O method proposed in this paper consistently achieves the lowest energy consumption, regardless of changes in the number of users. Although the energy consumption of all algorithms increases with the number of users, this is expected since, to meet the SINR requirements of more users, a higher transmission power is needed, leading to increased energy consumption. However, the projection-based L2O method exhibits the smallest rate of increase, and when the number of users exceeds four, the growth rate slows significantly. This suggests that the proposed L2O method has greater potential for large-scale user scenarios.

Fig.~\ref{fig-power-gamma} shows how energy consumption changes as the user's minimum SINR requirement, $\gamma$, increases. It is clear that the energy consumption of all algorithms, except Trust-Constr, rises as $\gamma$ increases. This is because a larger transmission power is required to maintain the desired SINR levels in the presence of increased noise. However, the projection-based L2O method consistently shows the lowest energy consumption across all values of $\gamma$, with the slowest growth rate. This reinforces the effectiveness and robustness of the proposed algorithm in handling varying SINR requirements.

\section{Conclusion}
In this paper, we have introduced a Hilbert projection-based learning-to-optimize (L2O) framework for constrained optimization (CO) in wireless networks and other engineering domains. By incorporating the projection operation as an activation function, the proposed method ensures strict feasibility of the solution space, thereby overcoming one of the primary limitations of conventional neural network (NN)-based approaches for CO. We further derived a heuristic gradient computation mechanism, enabling end-to-end, label-free training of NNs for both convex and non-convex problems under convex constraints. Simulation results demonstrated that the proposed method not only guarantees feasibility but also achieves a performance superior to or on par with existing algorithms, including traditional convex analysis-based techniques and penalty-based learning methods. Moreover, the ability to train in an unsupervised manner significantly enhances the scalability of our framework, making it suitable for a broad class of high-dimensional optimization tasks. Future work will explore extensions to mixed-integer programming, the integration of advanced NN architectures such as transformers and graph neural networks, and the application of our projection-based approach to real-time and large-scale wireless communication scenarios.

\bibliography{ref}

\begin{thebibliography}{10}
\providecommand{\url}[1]{#1}
\csname url@samestyle\endcsname
\providecommand{\newblock}{\relax}
\providecommand{\bibinfo}[2]{#2}
\providecommand{\BIBentrySTDinterwordspacing}{\spaceskip=0pt\relax}
\providecommand{\BIBentryALTinterwordstretchfactor}{4}
\providecommand{\BIBentryALTinterwordspacing}{\spaceskip=\fontdimen2\font plus
\BIBentryALTinterwordstretchfactor\fontdimen3\font minus \fontdimen4\font\relax}
\providecommand{\BIBforeignlanguage}[2]{{%
\expandafter\ifx\csname l@#1\endcsname\relax
\typeout{** WARNING: IEEEtran.bst: No hyphenation pattern has been}%
\typeout{** loaded for the language `#1'. Using the pattern for}%
\typeout{** the default language instead.}%
\else
\language=\csname l@#1\endcsname
\fi
#2}}
\providecommand{\BIBdecl}{\relax}
\BIBdecl

\bibitem{boyd2004convex}
S.~Boyd, ``Convex optimization,'' \emph{Cambridge UP}, 2004.

\bibitem{wang2022joint}
X.~Wang, L.~Fu, N.~Cheng, R.~Sun, T.~Luan, W.~Quan, and K.~Aldubaikhy, ``Joint flying relay location and routing optimization for 6g uav--iot networks: A graph neural network-based approach,'' \emph{Remote Sensing}, vol.~14, no.~17, p. 4377, 2022.

\bibitem{9252917}
Y.~Shen, Y.~Shi, J.~Zhang, and K.~B. Letaief, ``Graph neural networks for scalable radio resource management: Architecture design and theoretical analysis,'' \emph{IEEE Journal on Selected Areas in Communications}, vol.~39, no.~1, pp. 101--115, 2021.

\bibitem{de2002approximation}
E.~De~Klerk and D.~V. Pasechnik, ``Approximation of the stability number of a graph via copositive programming,'' \emph{SIAM Journal on Optimization}, vol.~12, no.~4, pp. 875--892, 2002.

\bibitem{sun2016majorization}
Y.~Sun, P.~Babu, and D.~P. Palomar, ``Majorization-minimization algorithms in signal processing, communications, and machine learning,'' \emph{IEEE Transactions on Signal Processing}, vol.~65, no.~3, pp. 794--816, 2016.

\bibitem{wen2024survey}
D.~Wen, Y.~Zhou, X.~Li, Y.~Shi, K.~Huang, and K.~B. Letaief, ``A survey on integrated sensing, communication, and computation,'' \emph{IEEE Communications Surveys \& Tutorials}, 2024.

\bibitem{an2005dc}
L.~T.~H. An and P.~D. Tao, ``The dc (difference of convex functions) programming and dca revisited with dc models of real world nonconvex optimization problems,'' \emph{Annals of operations research}, vol. 133, pp. 23--46, 2005.

\bibitem{niesen2007adaptive}
U.~Niesen, D.~Shah, and G.~Wornell, ``Adaptive alternating minimization algorithms,'' in \emph{2007 IEEE International Symposium on Information Theory}.\hskip 1em plus 0.5em minus 0.4em\relax IEEE, 2007, pp. 1641--1645.

\bibitem{nishihara2015general}
R.~Nishihara, L.~Lessard, B.~Recht, A.~Packard, and M.~Jordan, ``A general analysis of the convergence of admm,'' in \emph{International conference on machine learning}.\hskip 1em plus 0.5em minus 0.4em\relax PMLR, 2015, pp. 343--352.

\bibitem{8444648}
H.~Sun, X.~Chen, Q.~Shi, M.~Hong, X.~Fu, and N.~D. Sidiropoulos, ``Learning to optimize: Training deep neural networks for interference management,'' \emph{IEEE Transactions on Signal Processing}, vol.~66, no.~20, pp. 5438--5453, 2018.

\bibitem{8922744}
F.~Liang, C.~Shen, W.~Yu, and F.~Wu, ``Towards optimal power control via ensembling deep neural networks,'' \emph{IEEE Transactions on Communications}, vol.~68, no.~3, pp. 1760--1776, 2020.

\bibitem{goodfellow2016deep}
I.~Goodfellow, Y.~Bengio, and A.~Courville, \emph{Deep learning}.\hskip 1em plus 0.5em minus 0.4em\relax MIT press, 2016.

\bibitem{achiam2023gpt}
J.~Achiam, S.~Adler, S.~Agarwal, L.~Ahmad, I.~Akkaya, F.~L. Aleman, D.~Almeida, J.~Altenschmidt, S.~Altman, S.~Anadkat \emph{et~al.}, ``Gpt-4 technical report,'' \emph{arXiv preprint arXiv:2303.08774}, 2023.

\bibitem{bertsekas2003convex}
D.~Bertsekas, A.~Nedic, and A.~Ozdaglar, \emph{Convex analysis and optimization}.\hskip 1em plus 0.5em minus 0.4em\relax Athena Scientific, 2003, vol.~1.

\bibitem{borwein2006convex}
J.~Borwein and A.~Lewis, \emph{Convex Analysis}.\hskip 1em plus 0.5em minus 0.4em\relax Springer, 2006.

\bibitem{hong2015decomposition4}
M.~Hong, Q.~Li, and Y.-F. Liu, ``Decomposition by successive convex approximation: A unifying approach for linear transceiver design in heterogeneous networks,'' \emph{IEEE Transactions on Wireless Communications}, vol.~15, no.~2, pp. 1377--1392, 2015.

\bibitem{hunter2004tutorial5}
D.~R. Hunter and K.~Lange, ``A tutorial on mm algorithms,'' \emph{The American Statistician}, vol.~58, no.~1, pp. 30--37, 2004.

\bibitem{horst1999dc6}
R.~Horst and N.~V. Thoai, ``Dc programming: overview,'' \emph{Journal of Optimization Theory and Applications}, vol. 103, pp. 1--43, 1999.

\bibitem{an2005dc7}
L.~T.~H. An and P.~D. Tao, ``The dc (difference of convex functions) programming and dca revisited with dc models of real world nonconvex optimization problems,'' \emph{Annals of operations research}, vol. 133, pp. 23--46, 2005.

\bibitem{cai2017convergence8}
X.~Cai, D.~Han, and X.~Yuan, ``On the convergence of the direct extension of admm for three-block separable convex minimization models with one strongly convex function,'' \emph{Computational Optimization and Applications}, vol.~66, pp. 39--73, 2017.

\bibitem{nishihara2015general9}
R.~Nishihara, L.~Lessard, B.~Recht, A.~Packard, and M.~Jordan, ``A general analysis of the convergence of admm,'' in \emph{International conference on machine learning}.\hskip 1em plus 0.5em minus 0.4em\relax PMLR, 2015, pp. 343--352.

\bibitem{deng2016global10}
W.~Deng and W.~Yin, ``On the global and linear convergence of the generalized alternating direction method of multipliers,'' \emph{Journal of Scientific Computing}, vol.~66, pp. 889--916, 2016.

\bibitem{wang2014convergence11}
F.~Wang, Z.~Xu, and H.-K. Xu, ``Convergence of bregman alternating direction method with multipliers for nonconvex composite problems,'' \emph{arXiv preprint arXiv:1410.8625}, 2014.

\bibitem{chen2016direct12}
C.~Chen, B.~He, Y.~Ye, and X.~Yuan, ``The direct extension of admm for multi-block convex minimization problems is not necessarily convergent,'' \emph{Mathematical Programming}, vol. 155, no.~1, pp. 57--79, 2016.

\bibitem{zhang2020proximal}
J.~Zhang and Z.-Q. Luo, ``A proximal alternating direction method of multiplier for linearly constrained nonconvex minimization,'' \emph{SIAM Journal on Optimization}, vol.~30, no.~3, pp. 2272--2302, 2020.

\bibitem{hopfeild1985neural13}
J.~Hopfeild and D.~Tank, ``Neural computation of decision in optimization problems,'' \emph{Biological cybernetic}, pp. 52--60, 1985.

\bibitem{bian2009subgradient14}
W.~Bian and X.~Xue, ``Subgradient-based neural networks for nonsmooth nonconvex optimization problems,'' \emph{IEEE Transactions on Neural Networks}, vol.~20, no.~6, pp. 1024--1038, 2009.

\bibitem{gao2009new15}
X.-B. Gao and L.-Z. Liao, ``A new projection-based neural network for constrained variational inequalities,'' \emph{IEEE transactions on neural networks}, vol.~20, no.~3, pp. 373--388, 2009.

\bibitem{cheng2019end16}
R.~Cheng, G.~Orosz, R.~M. Murray, and J.~W. Burdick, ``End-to-end safe reinforcement learning through barrier functions for safety-critical continuous control tasks,'' in \emph{Proceedings of the AAAI conference on artificial intelligence}, vol.~33, no.~01, 2019, pp. 3387--3395.

\bibitem{pan2020deepopf17}
X.~Pan, T.~Zhao, M.~Chen, and S.~Zhang, ``Deepopf: A deep neural network approach for security-constrained dc optimal power flow,'' \emph{IEEE Transactions on Power Systems}, vol.~36, no.~3, pp. 1725--1735, 2020.

\bibitem{zamzam2020learning18}
A.~S. Zamzam and K.~Baker, ``Learning optimal solutions for extremely fast ac optimal power flow,'' in \emph{2020 IEEE international conference on communications, control, and computing technologies for smart grids (SmartGridComm)}.\hskip 1em plus 0.5em minus 0.4em\relax IEEE, 2020, pp. 1--6.

\bibitem{kotary2021end19}
J.~Kotary, F.~Fioretto, P.~Van~Hentenryck, and B.~Wilder, ``End-to-end constrained optimization learning: A survey,'' \emph{arXiv preprint arXiv:2103.16378}, 2021.

\bibitem{nellikkath2021physics20}
R.~Nellikkath and S.~Chatzivasileiadis, ``Physics-informed neural networks for minimising worst-case violations in dc optimal power flow,'' in \emph{2021 IEEE International Conference on Communications, Control, and Computing Technologies for Smart Grids (SmartGridComm)}.\hskip 1em plus 0.5em minus 0.4em\relax IEEE, 2021, pp. 419--424.

\bibitem{donti2021dc321}
P.~L. Donti, D.~Rolnick, and J.~Z. Kolter, ``Dc3: A learning method for optimization with hard constraints,'' \emph{arXiv preprint arXiv:2104.12225}, 2021.

\bibitem{grant2014cvx}
M.~Grant and S.~Boyd, ``{CVX}: Matlab software for disciplined convex programming, version 2.1,'' 2014.

\end{thebibliography}
\bibliographystyle{IEEEtran}
\ifCLASSOPTIONcaptionsoff
  \newpage
\fi

\end{document}